# Reconsidering Optimistic Algorithms for Relational DBMS

Malcolm Crowe

School of Computing, Engineering and Physical Sciences
University of the West of Scotland
Paisley, UK
e-mail: malcolm.crowe@uws.ac.uk

Fritz Laux

Department of Informatics
Reutlingen University
Reutlingen, Germany
e-mail: fritz.laux@fh-reutlingen.de

*Abstract*—**At DBKDA 2019, we demonstrated that StrongDBMS with simple but rigorous optimistic algorithms, provides better performance in situations of high concurrency than major commercial database management systems (DBMS). The demonstration was convincing but the reasons for its success were not fully analysed. There is a brief account of the results below. In this short contribution, we wish to discuss the reasons for the results. The analysis leads to a strong criticism of all DBMS algorithms based on locking, and based on these results, it is not fanciful to suggest that it is time to re-engineer existing DBMS.**

## I. INTRODUCTION

While the Standard Query Language (SQL) standard [9] famously describes the well-known four transaction levels of read uncommitted, read committed, repeatable read, and serializable, it wisely does not mandate any particular strategy for ensuring correct transaction behaviour, as explained in Note 47 [9]. However, all commercial database management systems (DBMS) use locking to ensure correct transactional behaviour in the face of concurrent accesses to a database.

This approach, with the attendant use of pessimistic concurrency algorithms, may have seemed attractive in 1974, and is still the easiest to explain. If the client has acquired locks on all the data it needs, it appears that a successful commit can be guaranteed. However, if the client and server are communicating over a network, the Consistency-Availability-Partition tolerance (CAP) theorem and the two-army thought experiment both demonstrate that the success of the commit may be indefinitely delayed unless the client's locks are overridden. To these theoretical objections two practical considerations can be added, first, that locking systems are complex, so that deadlocks are almost unavoidable, and, second, that client-side locks are subject to timeout. As a result, the apparent guarantee of success does not work well over the internet where interactive clients expect to have a comparatively long time to complete a transaction.

In practice, many software developers instead use application-level protocols to provide optimistic concurrency for distributed applications communicating with web-servers that handle all access to the database. The resulting mismatch of concurrency strategies between application and database has led to middleware trying to provide a concurrency mechanism that is more application affine and abstract from the database provided concurrency control (e.g., see [2, 3, 14]). But, far from solving the problem of transaction coordination, this only compounds the problem by adding another competing source of persistence, and the difference in approach to concurrency does not help. It becomes natural to ask whether the database server itself should also use optimistic algorithms for concurrency control

The significance of the StrongDBMS [7] demonstration [1][15] was that its optimistic algorithms were extremely simple and startlingly effective in providing fully serializable transactions under conditions of high data conflict. The experiment was set up so that correct operation would necessitate most transactions failing to commit, but much greater overall throughput resulted from StrongDBMS' optimistic operation. StrongDBMS's transaction log demonstrated that all committed transactions had been serialised, despite the large number of overlapping long transactions.

The implementation of StrongDBMS was also interesting in featuring the use of immutable data structures, and it seems plausible that all the usual DBMS features could be implemented using this approach. Work has been progressing since DBKDA 2019 to achieve this by modifying the existing PyrrhoDB to use a similar architecture to StrongDBMS.

The structure of this paper is as follows. Section 2 contains an analysis of the reasons for the demonstrated differences in performance between optimistic DBMS (such as StrongDBMS and PyrrhoDB) and other systems. Section 3 explains some minor departures from standard SQL semantics in the test. Section 4 discusses the details of the modified benchmark test used in the demonstration. Section 5 presents a synopsis of the test results. Finally, Section 6 summarises the conclusions of this study.

## II. CONFLICT DETECTION AND ROLLBACK

The essential point of optimistic transactions is that conflicts are detected only at the end of the transaction when commit is attempted. At this point, if it is found that conflicts have occurred, the commit will fail, and none of the transaction's work will be written to the database.

This approach is sometimes called First Committer Wins (FCW). It has the advantage that short transactions are more likely to succeed. In the literature [4, page 170], it has been assumed that FCW systems would have high validation costs or reduced throughput because of unnecessary rollbacks that would occur if the check includes only 'dangerous structures' [5]. But the demonstration showed that, when combined with optimistic execution, throughput was enhanced through use of FCW. Some database textbooks suggest that optimistic execution is inherently less effective than the usual locking-based approach when load is high, but this is now seen to be another myth. In the rare situation where transactions access the same data (hot spot), it might be possible that a transaction is repeatedly aborted (starving problem).

## III. SOME DEROGATIONS

For simplicity, we focus exclusively on SERIALIZABLE transactions. It seems worthwhile here to explain other technical respects in which the implementations depart from the standard description. The standard stipulates that all changes made on commit are accessible to concurrent transactions. We interpret this as excluding concurrent *serializable* transactions, as it is more natural that a serialisable transaction continues to see the database as it stood at the time the transaction started ("snapshot isolation"), apart from the changes it is making. In the case that the transaction does not intend to commit changes, it is intrusive to advise on changes that other users have made.

It is well known that snapshot isolation is insufficient to ensure consistency [6]. Even optimistic algorithms need to lock the database *during commit* while the transaction is checked for conflicts. This however is quite different from acquiring locks at an earlier stage in the transaction.

One further simplification in our work is always to enforce constraints and integrity checks . For example, the "no action" options are disallowed for referential constraints. This ensures that the database is kept in a consistent state even after each step in a schedule. For constraints that cannot be satisfied with one SQL-statement, our chosen solution is to allow deferral of triggers to the end of a transaction.

## IV. THE CASE STUDY

The demonstration of StrongDBMS used the Transaction Processing Council Benchmark C (TPC-C) [13] with a modification to create high levels of data conflict between clerks who enter new orders for a warehouse.

To begin with, the TPC-C benchmark normally has 1 clerk per warehouse, so that the conflict rate is around 4%. In the reported tests, we deliberately increased the concurrency challenge by using multiple clerks for a single warehouse. When the number of clerks goes above 10, most New Order tasks will fail with a write-write conflict on the next order number for the district (NEXT_O_ID) as there are only 10 districts. Worse, the single row in the WAREHOUSE table contains a running total for the year (W_YTD), which is updated by the payment task, and fields from this row are read by all the NewOrder tasks and others so that a great many more tasks are aborted because of read/write conflicts. In all the products tested, apart from PyrrhoDB and StrongDBMS, read/write conflicts are detected at the row level or wider.

Both PyrrhoDB and StrongDBMS see no conflict between the Payment and NewOrder task because Payment is the only task that accesses W_YTD, and one of the available tests in the ReadConstraint for detecting read/write conflicts is a set of fields in a specific single row of a table.

There are actually three levels of read/write conflict detection in these DBMS. The following comment in the source code for Read Set dates from about 2005 [8] (tb refers to the base table affected):

"ReadConstraints record all of the objects that have been accessed in the current transaction so that this transaction will conflict with a transaction that changes any of them. However, for records in a table, we allow specific non-conflicting updates, as follows:

"(a) (CheckUpdate) If unique selection of specific records cannot be guaranteed, then we should report conflict if any column read is updated by another transaction.

"(b) (CheckSpecific) If we are sure the transaction has seen a small number of records of tb, selected by specific values of the primary or other unique key, then we can limit the conflict check to updates of the selected records (if any), or to updates of the key TableColumns.

"(c) (BlockUpdate) as (a) but it is known that case (b) cannot apply."

If the isolation level is reduced to repeatable-read or read-committed, most of the competing products achieve performance comparable with Pyrrho and StrongDBMS. However, there is a risk that the database may show wrong results or an inconsistent state. This is what we found for a commercial product.

The use of escrow methods [11][12] could avoid hot spot conflicts like in NEXT_O_ID (resp. W_YTD) for many DBMS if the semantics is known, e.g., an increment semantic (resp. commutative semantics). Laiho and Laux [10] also developed a method of using row-versioning to ensure correct non-blocking operation of distributed applications. Both these approaches require changes to the application protocols, but they can be used with existing commercial DBMS products.

## V. THE BENCHMARK RESULTS

The TPC-C benchmark simulates a telephone-based order entry system for 100000 products where each warehouse has 30000 customers assigned to 10 districts. There is one clerk per warehouse, and the simulation includes a randomised set of tasks with time-delays so that a realistic work rate for the clerk is simulated, allowing the clerk to process 16 new orders in 10 minutes: each order has between 5 and 15 lines. There is some scope for concurrency verification for the DBMS, as items can be supplied from other warehouses, and the specification results in about 4% of conflicting transactions.

We adapted this test by providing multiple clerks for a single warehouse, and then the database design results in much higher levels of conflict as described above. In the 10-minute experiments, the maximum number of new orders per clerk remains 16, but the actual throughput will be much less owing to transaction conflict. DBMS generally allow a range of transaction isolation levels. From the viewpoint of this paper, the interesting results are for SERIALIZABLE transactions only.

The initial state of the database, and the details of what the tasks involve, are specified in great detail on the TPC-C website. In simple terms, each task requires committing some changes to the database. Many of the tasks perform a single insert or update on a single table. The commit for the new order task inserts new rows in HISTORY, ORDER and ORDER_LINE (5 to 15 order lines per order) and updates WAREHOUSE, DISTRICT, CUSTOMER and 5 to 15 rows in STOCK. All the updates involved in a new order have a good chance of conflict since there is only 1 warehouse and 10 districts. There is a smaller chance of conflict on STOCK and CUSTOMER since there are more of these. The distinction between ORDER and NEW_ORDER is that customers are expected to pay for completed ORDERS, and NEW_ORDERS require delivery. In the 10 minute test, the delivery for a NEW_ORDER might be scheduled but won't complete.

For StrongDBMS, we found the behaviour shown in Table I. This shows 241 (= 30241 - 30000) new orders for 30 clerks, and also indicates the reported number of failed transactions (="Exceptions").

TABLE I.    RESULTS FOR STRONGDBMS

| Name | Initial | 1 clerk | 10 clerks | 20 clerks | 30 clerks |
|---|---|---|---|---|---|
| Commits | 0 | 39 | 302 | 512 | 565 |
| Exceptions | 0 | 0 | 104 | 387 | 1071 |
| ORDER | 30000 | 30016 | 30138 | 30199 | 30241 |
| NEW_ORDER | 9000 | 9016 | 9138 | 9199 | 9241 |
| ORDER_LINE | 285007 | 285158 | 286207 | 286638 | 286965 |
| DELIVERY | 0 | 1 | 13 | 22 | 32 |

A major commercial DBMS, using serializable transaction isolation, completed only 132 NEW_ORDERS for 30 clerks, as shown in Table II.

TABLE II.    RESULTS FOR COMMERCIAL DEBMS (SERIALIZABLE, USING 2PL)

| Name | Initial | 1 clerk | 10 clerks | 20 clerks | 30 clerks |
|---|---|---|---|---|---|
| Commits | 0 | 41 | 211 | 276 | 290 |
| Exceptions | 0 | 0 | 43 | 132 | 213 |
| ORDER | 30000 | 30016 | 30111 | 30127 | 30132 |
| NEW_ORDER | 9000 | 9016 | 9111 | 9127 | 9132 |
| ORDER_LINE | 285007 | 285158 | 286114 | 286223 | 286295 |
| DELIVERY | 0 | 1 | 12 | 18 | 18 |

The commercial DBMS frequently aborted the transaction with a report of deadlock, without attempting to commit.

Some investigation took place on using other isolation levels and other DBMS. These tests are reproducible, and versions of the software for several major commercial DBMS are available on the GitHub website [16]. However, this software is implemented with a thread for each clerk with its own database connection, and in some cases this seemed to result in the DBMS erroneously reporting that transactions were being nested, or already completed.

Callum Fyffe continued the tests for StrongDBMS to over 100 clerks [14], and while the numbers continued to rise, eventually the results became less reproducible as the operating system intervened to deal with memory saturation. Our collected results for SERIALIZABLE isolation are shown in Table III, where the asterisks indicate that further tests were not carried out owing to reducing throughput.

TABLE III.    FURTHER RESULTS

| Name | 1 clerk | 2 | 5 | 10 | 20 | 30 | 40 | 50 | 60 |
|---|---|---|---|---|---|---|---|---|---|
| StrongDBMS laptop | 16 | | | 138 | 199 | 241 | * | | |
| StrongDBMS 16GB RAM | 16 | | | 129 | 220 | 254 | 409 | 331 | 328 |
| Commercial 1 | 16 | | | 111 | 127 | 132 | 16 | * | |
| Commercial2 | 16 | | | 107 | 114 | 119 | 124 | 117 | * |
| Commercial3 | 16 | 33 | 69 | 6 | * | | | | |

Figure 1 shows the comparable results from Table III as a chart.

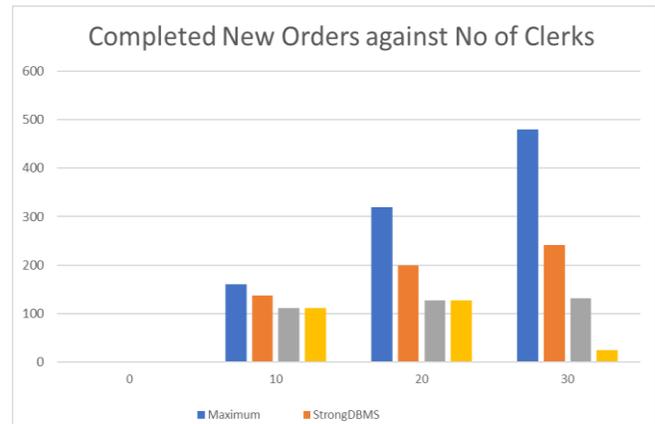

Figure 1. Comparable test results. The first bar in each group shows the maximum possible (16x number of clerks), and the second is StrongDBMS.

## VI.    CONCLUSIONS

The study reported here makes a case for extending optimistic algorithms to other database products. This would provide a radical and welcome way of removing the "impedance mismatch" between application and DBMS protocols. Myths about such algorithms are deeply entrenched in the database community, but it is time for better and more considered analysis.